\documentclass[11pt,a4paper]{article}
\usepackage{jcappub}

\def\be{\begin{equation}}
\def\ee{\end{equation}}
\def\bea{\begin{eqnarray}}
\def\eea{\end{eqnarray}}

\begin{document}

\title{Cosmic Logic: a Computational Model}

\author{Vitaly Vanchurin}

\emailAdd{vvanchur@d.umn.edu}

\date{\today}

\affiliation{Department of Physics and Astronomy, University of Minnesota, Duluth, Minnesota, 55812}

\abstract{ 
We initiate a formal study of logical inferences in context of the measure problem in cosmology or what we call {\it cosmic logic}. We describe a simple computational model of cosmic logic suitable for analysis of, for example, discretized cosmological systems. The construction is based on a particular model of computation, developed by Alan Turing, with cosmic observers (CO), cosmic measures (CM) and cosmic symmetries (CS) described by Turing machines. CO machines always start with a blank tape and CM machines take CO's Turing number  (also known as description number or G{\" o}del number) as input and output the corresponding probability. Similarly, CS machines take CO's Turing number as input, but output either one if the CO machines are in the same equivalence class or zero otherwise. We argue that CS machines are more fundamental than CM machines and, thus, should be used as building blocks in constructing CM machines. We prove the non-computability of a CS machine which discriminates between two classes of CO machines: {\it mortal} that halts in finite time and {\it immortal} that runs forever. In context of eternal inflation this result implies that it is impossible to construct CM machines to compute probabilities on the set of all CO machines using cut-off prescriptions.  The cut-off measures can still be used if the set is reduced to include only machines which halt after a finite and predetermined number of steps. 
}

\maketitle

\section{Introduction}

Measure problem in cosmology is formulated as a problem of assigning probabilities to cosmic observations/observers. Although the problem is very general it is usually studied in context of eternal inflation \cite{EternalInflation, EternalInflation2, EternalInflation3}. The statement of the problem assumes that the set of possible observations/observers, ${\cal O}$, was already identified and the remaining task is to define a probability measure on that set, $\mu: {\cal O} \rightarrow [0,1]$. However, in reality even the task of defining  ${\cal O}$ is ambiguous. For example, it remains unclear whether the relevant observations are localized events (as is often assumed \cite{Vilenkin}) or if they can be extended (as was discussed in \cite{GuthVanchurin}) or even infinite-size (as was considered in \cite{Vanchurin}) objects. Note that at this stage we are not concerned with probabilities of the individual observations, but only with construction of a set of all such  events. Only after the set ${\cal O}$ is constructed one can study the different probability measures on  ${\cal O}$ in an attempt to identify the correct one. 

However, given a large number of choices for $\mu$ one might wonder whether the problem is well posed even if the set ${\cal O}$ was identified. So far the two main guiding principles in selecting probability measures were the operational simplicity (i.e. easy to define and compute) and phenomenological validity (i.e. agreement with observations). This is not very satisfactory and one might hope to derive $\mu$ from first principles (see for example Refs. \cite{Vanchurin} and \cite{Continuum} for our recent attempts to tackle the measure problem from this perspective). In our studies we shall adopt a view that this should be possible and propose that the guiding principle should be symmetries of the set ${\cal O}$ instead of simplicity or phenomenology. If we denote the group of symmetry transformations by ${\cal S}$, then the invariance of probability measures would imply $\mu(s(o)) = \mu(o)$ for all $o \in {\cal O}$ and $s \in {\cal S}$ (See for example Refs. \cite{Continuum} and \cite{Tree} for discussion of symmetries in some toy models of eternal inflation).  

Of course, given a set of observational events $\cal O$ and a group of symmetries ${\cal S}$ it is still unclear whether there exist a unique and invariant measure $\mu$, but it is very tempting to explore this possibility based on a fundamental result from mathematical analysis. According to Haar's theorem there exist is a unique measure on subsets of a locally compact group which is invariant under actions of the group \cite{Haar}. The most prominent example of such a measure is the Lebesgue measure on a set of real numbers which is a unique measure invariant under translations. (See Ref. \cite{Continuum} for a recent discussion of the Lebesgue measure in some toy models of eternal inflation). In other words one might want to conjecture that it should be possible to construct a unique probability measure $\mu$ which is invariant under all of the symmetries of ${\cal O}$ once the group of symmetries ${\cal S}$ is constructed. If correct then the problem of finding the correct probability measure $\mu$ on $\cal O$ may be replace with a problem of defining a symmetry group ${\cal S}$ on $\cal O$. Conversely, one can use the symmetry considerations to rule out certain probability measures as will be demonstrated in this paper with respect to the so-called cut-off measures.

So far the treatment of the triplet $({\cal O}, {\cal S}, \mu)$ was very general, but our final goal is to be able to make logical inferences about cosmic observations/observers. This means that the construction of the triplet must be based on certain logical system (we call {\it cosmic logic}) that would be useful for making cosmological predictions. At this point it remains unclear what should be the formal system of cosmic logic, but there are certain aspects of the possible logical inferences that had already been analyzed in the literature. In particular it was shown that the probabilistic logic applied to the models of eternal inflation often leads to paradoxes (e.g. Guth-Vanchurin paradox \cite{GuthVanchurin, GuthVanchurin2, GuthVanchurin3, GuthVanchurin4}, Youngness paradox \cite{Youngness, Youngness2, Youngness3} and Boltzmann brains problem \cite{Boltzmann, Boltzmann2, Boltzmann3}). This might suggest that the stochastic model of eternal inflation is incomplete and requires a major revision, or  that the methods for extracting probabilities based on the often-used cut-off prescriptions are not appropriate for cosmological systems. In either case it is impossible to identify the problem until we have a formal logical system for making cosmological prediction (i.e. cosmic logic). 

In this paper we discuss a simple computational model of cosmic logic which gives a precise operational meaning to the triplet  $({\cal O}, {\cal S}, \mu)$  by modeling all three elements as Turing machines. Turing machine is a simple hypothetical device introduced by Alan Turing in 1936 \cite{Halting}. Despite of the apparent simplicity it is capable of simulating the logic of arbitrary computer algorithm. In this paper the Turing machines will be used for computations in three different contexts. The first type of computation is identified with cosmic observers/observations ${\cal O}$ (we denote by CO), the second type with cosmic measures $\mu$ (we denote by CM) and the third type with cosmic symmetries ${\cal S}$ (we denote by CS).

The paper is organized as follows. In Sec. \ref{Sec:TuringMachines} we give a pedagogical introduction to Turing's model of computation and also discuss a problem of describing relativistic observers using Turing machines. In Sec. \ref{Sec:CosmicObservers} we define cosmic observer (CO) machines whose Turing number once inputed into cosmic measure (CM) machine would output its probability. In Sec. \ref{Sec:Symmetries}  we discuss cosmological symmetry (CS) machines and use them to prove the non-exsitance  of CM machines corresponding to symmetries shared by all (global and local) cut-off measures. In Sec. \ref{Sec:Discussion} we summaries and discuss the main results of the paper.

\section{Turing Machines}\label{Sec:TuringMachines}

The ultimate goal of cosmic logic is to enable cosmic observers to make logical inferences about universe. In particular observers should be able to check weather a given logical statement about universe is true, false or, more generally, true with some probability $p$. But what does it mean to check whether a give logical statement is true? In modern terminology it means that we want a program which would take the statement as input and output $0$ if it is false, $1$ if it is true or $p$ if the statement is true with probability $p$. Programs are to be written in some programing language (software) and on some computer (hardware). So, in order to make the ``check of statements'' precise we have to define some software/hardware combination which would be capable of preforming desired computational tasks. Note that we are not interested in building an actual computer, but to have a simple hypothetical model to formalize the task of logical inference. 

Turing Machine does exactly that. There are many (computationally equivalent) variants of Turing machines, but we will stick with a simple machine which consists of only three major components: TAPE, REGISTER and PROGRAM. There are three analogies that might help one to grasp the concept of Turing machines. It is might be useful to think of a Turing machine as a person performing calculations (as was originally described by Turing \cite{Halting}), as a computer (as is often described in modern books) or as a relativistic observer (as will be explained in the text below). For the person performing calculation the TAPE may be nothing but a notebook,  state of REGISTER may be the person's ``state of mind'' and PROGRAM is the person's act of performing calculations based on the current state of mind and what is written on (the current page of) the notebook. From the point of view of modern computers, the TAPE is computer memory,  state of the REGISTER is a state of processor and the PROGRAM is a software.

What is common in both examples is that there is a clear distinction between state of the REGISTER and state of the TAPE. However, for physical systems there is usually one type of state whose dynamics is governed by some equations of motion (which would be described by the PROGRAM in the case of Turing machines.) Thus to draw analogies between physical systems and Turing machines we need to divide the state space into internal states describing the REGISTER and external state describing the TAPE. What is important to note is that the state space of the REGISTER is finite, but the state space of the TAPE is infinite (although only a finite portion of it may be used at any given time). Moreover, when the PROGRAM is executed the entire state of the REGISTER is participating in dynamics, but situation is very different for the TAPE. There only the state of a single cell (i.e. current cell) is affected by the PROGRAM and all other cells remain unaffected.

Consider a relativistic observer (e.g. a rocket, a plant, a solar system, etc.) on the background of curved manifold. Such observer can usually be modeled with a finite number of coarse-grained degrees of freedom whose dynamics is only affected by the state of the local degrees of freedom of geometry. Moreover the system can back-react on the geometry and change its local state. Then we could associate the internal degrees of freedom of the observer with REGISTER, external degrees of freedom of the geometry with TAPE and the evolution laws for the local system with PROGRAM. The read-write head would always points to current position on (in this case three dimensional) TAPE which corresponds to the current position of the observer on  manifold.  Despite of many similarities there is an important difference between the usual Turing machine and relativistic systems discussed here. The local changes in the geometry can causally propagate to other location on manifold away from the position of the observer (e.g. gravitational waves), but the changes of a current cell on the TAPE do not propagate to other cells. We will come back to this issue after cosmic observers are introduced in Sec. \ref{Sec:CosmicObservers}.

\begin{figure}
\begin{center}
\includegraphics[width=0.75\textwidth] {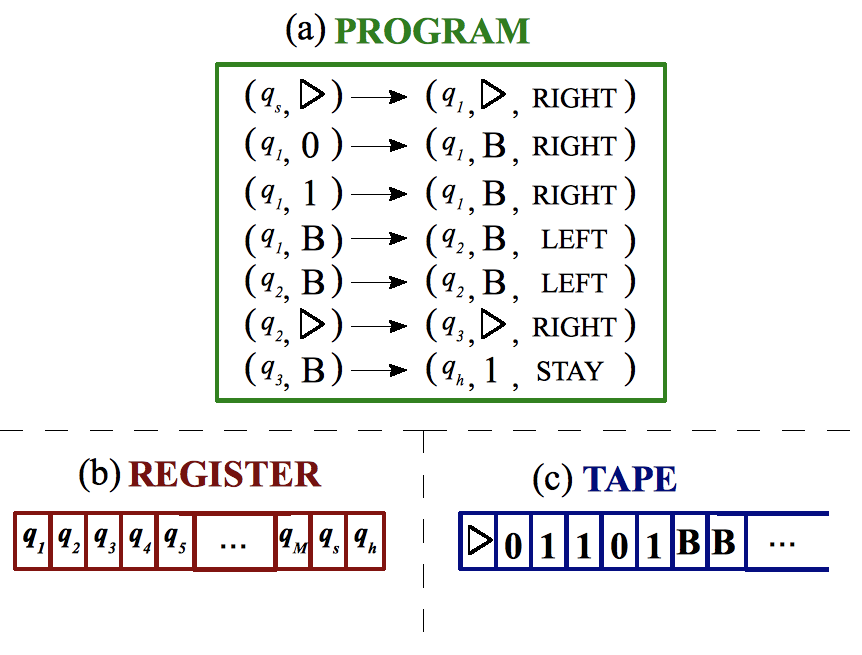}
\caption{Three components of a Turing machine: (a) PROGRAM, (b) REGISTER and (c) TAPE.  \label{plot}}
\end{center}
\end{figure}

We can now give a formal definition of the three components of Turing machine:
\begin{itemize}
\item {\bf TAPE} is a semi-infinite collections of symbols from alphabet 
\be
\Gamma=\{\vartriangleright,\text{B},0,1\}.
\ee
The edge symbol $\vartriangleright$ is occupied by only leftmost cell, symbols $0$ and $1$ are occupied by only a finite number of cells, and all of the remaining cells are occupied by blank symbol B (See Fig. \ref{plot}.c). The symbols are passed to and from the TAPE using  {\it read-write head}. At any moment the  read-write head points to only a single cell (current position) from where a symbol can be read or to where a symbol can be (over)written. 

\item {\bf REGISTER} contains a finite set of control states
\be
Q=\{q_{1},q_{2}...q_{M},q_{s},q_{h}\}
\ee
with two special states $q_{s}$ starting state and $q_{h}$ halting state (See Fig. \ref{plot}.b). Although the total number of states (i.e. $M+2$ ) may be large it is finite in contrast to the total number of states of infinite TAPE which is (uncountably) infinite. 

\item {\bf PROGRAM} is a list of instructions describing a map
\be
\delta:\Gamma \times Q \rightarrow \Gamma\times Q \times\{\text{LEFT} , \text{RIGHT}, \text{STAY}\}.
\ee
(See Fig. \ref{plot}.a). Depending on the state of a current cell on the TAPE and a state of the REGISTER the PROGRAM changes state of the current cell on the TAPE to a new state and state of the REGISTER to a new state. Then the PROGRAM also moves the read-write head to the left (LEFT), to the right (RIGHT) or does not move at all (STAY). 
\end{itemize}

Thus formally Turing machine is a triplet  $(\Gamma,Q,\delta)$. For example, a Turing machine which overwrites whatever was written on the TAPE with $1$ is given by the following PROGRAM
\begin{align}
& \delta(q_{s},\vartriangleright)  = (q_{1},\vartriangleright, \text{RIGHT})\notag \\
& \delta(q_{1},0)  = (q_{1},\text{B},\text{RIGHT})\notag \\
& \delta(q_{1},1)  = (q_{1},\text{B},\text{RIGHT})\notag \\
& \delta(q_{1},\text{B})  = (q_{2},\text{B},\text{LEFT})\notag \\
& \delta(q_{2},\text{B})  = (q_{2},\text{B},\text{LEFT})\notag \\
& \delta(q_{2},\vartriangleright)  = (q_{3},\vartriangleright,\text{RIGHT})\notag \\
& \delta(q_{3},\text{B})  = (q_{h},1,\text{STAY}). \label{Eq:Example}
\end{align}
The three components of the machine are also illustrated on Fig. \ref{plot}
 where the TAPE is in some arbitrary initial state. It is recommended that the reader (mentally) runs the program to confirm that it does what was prescribed, i.e. evaluates  function  M$(x)=1$.\footnote{We slightly abuse the notations here and throughout the paper by using the same symbol for the Turing machine and function that it evaluates (e.g. M and M$(x)$).} 

Since every Turing machine is completely specified by the triplet $(\Gamma,Q,\delta)$ which can be written as data in binary code, the binary code uniquely determines the machine and defines the so called Turing number (or description number or G{\" o}del number) that we denote with angular brakets, i.e. $\langle$M$\rangle$ . Note that the Turing number $\langle$M$\rangle$ of the machine M is essentially the algorithmic (or Kolmogorov) complexity of the output of M when the input of the machine is a blank TAPE. For example, analysis of fundamental constant in terms of their Kolmogorov complexities was carried in Ref. \cite{NumericalSearch}. 

The above example \eqref{Eq:Example} calculates a very simple function, M$(x)=1$, but it turns out that Turing machines can preform essentially all operations (computations) that can be programmed on modern computers. Such a universality of Turing Machine is sometimes referred to as Church-Turing thesis: \emph{The class of functions computable by a Turing machine corresponds exactly to the class of functions which we would normally regard as being computable by an algorithm.}  More generally one might consider the so-called Physical Church-Turing thesis: \emph{All physically computable functions are computable by Turing machines.} In this paper we propose a Cosmological Church-Turing thesis: \emph{All cosmic measures are computable by Turing machines.}

\section{Cosmic Observers}\label{Sec:CosmicObservers}

\begin{figure}
\begin{center}
\includegraphics[width=0.4\textwidth] {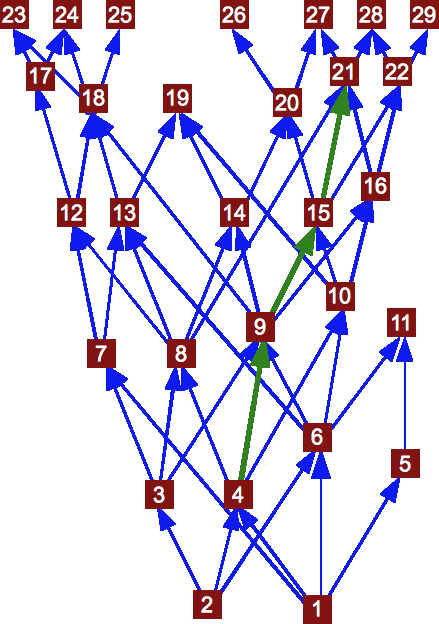}
\caption{Directed graph of a discretized universe. \label{plot2}}
\end{center}
\end{figure}

In Appendix \ref{Sec:DiscreteUniverse} we describe a discretization procedure which allows us to treat the entire universe as a directed graph with vertices $\{V_1, V_2, ...\} $ connected in chronological order by edges (whenever $E(V_i, V_j)=1$) and with fields  $\{\Phi_1, \Phi_2, .... \}$ taking only discrete values on these vertices. The procedure is certainly not unique but in some sense representative of its continuum counterpart. Construction can be thought of as a coordinate independent generalization of the (mesh refinement) construction of Ref. \cite{GuthVanchurin}. A sample discretized universe is illustrated on Fig. \ref{plot2}
 with red squares representing vertices and arrows representing directed edges.

By viewing the universe as a graph we can describe cosmic observations as paths (finite or infinite) connecting vertices in chronological order. Then an observation can be defined as a map 
\be
\alpha: \{1,...,N\} \rightarrow \mathbb{N}\label{Eq:o}
\ee
satisfying the chronology condition
\be
E(V_{\alpha(i)}, V_{\alpha(i+1)})=1\;\;\;\forall\;\;\; i\in\{1,...,N\}.
\ee
where if $N = \infty$ then $\{1, ..., N\} = \mathbb{N}$.\footnote{Note that for finite $N < \infty$ there might still exist $i\in\mathbb{N}$ for which $E(V_{\alpha(N)}, V_i)=1$ and thus, the observer must not always terminate in singularities.} A sample cosmic observation is illustrated on Fig. \ref{plot2} with a path connecting vertices along green arrows ($4 \rightarrow 9$, $9 \rightarrow 15$ and $15\rightarrow 21$), and thus $\alpha(1)=4, \alpha(2)=9, \alpha(3)=15$ and $\alpha(4)=21$. 

The map of Eq. \eqref{Eq:o}  defines a discretized world-line of some hypothetical observer who would at most be able to observe a sequence of numbers $\{\Phi_{\alpha(1)},..., \Phi_{\alpha(N)}\}$ describing field values encountered along the word-line. If we now imagine these numbers to be written on a TAPE, then one can ask what should be a Turing machine whose time evolution would generate the sequence $\{\Phi_{\alpha(1)},..., \Phi_{\alpha(N)}\}$ starting from a blank TAPE. We will refer to such Turing machine as a cosmic observer (CO) machine, i.e.
\be
\text{CO} \in {\cal O} \equiv \{ \text{Turing machines with initially blank TAPE} \}.
\ee
If $N$ is finite, then the {\it mortal} CO machine halts in $N$ steps and if $N$ is infinite then the {\it immortal} CO machine would never halt. 

This idea is not completely new as there were already attempts to identify observers with computers, but without relying on a particular model of computation \cite{ComputerObserver}. In this paper we will restrict our attention to Turing's model of computation and we will also allow more general observers/observations with both finite and infinite temporal extends (i.e. mortal and immortal CO machines respectively). The above description of CO machines was based on discretized values of fields which might be too precise for the purpose of making cosmological predictions. More generally (and also practically), what is actually written on the TAPE could be any fraction of information which is internal to the observer. For example, it might be more useful to think of the information on the TAPE as the actual knowledge of the cosmic observer in question about universe.

We can now come back to the issue briefly mentioned in Sec. \ref{Sec:TuringMachines}. It was argued that if one attempts to model relativistic observers as Turing machines then it is necessary to split the state into  degrees of freedom representing the REGISTER and those representing the TAPE. The most obvious choice was to treat the manifold as a TAPE and the internal degrees of freedom of the observer as a REGISTER. However, this leads to a problem that cells on the TAPE are constantly updated away from the read-write head (away from the current location of observer). To avoid this difficulty we propose to consider an alternative description of cosmic observers (CO) introduced above where the roles of internal and external degrees of freedom are switched (as compared to relativistic observer of Sec. \ref{Sec:TuringMachines}). The internal degrees of freedom are now associated with the TAPE and the external degrees of freedom are associated with REGISTER whose state determines what is written and over-written on the TAPE. The PROGRAM as before describes evolution laws which determine how the states of the REGISTER and TAPE evolve and how the information about these states is written on the TAPE. This view also suggests that the total number of external to observer degrees of freedom should be finite (finite number of states of REGISTER), but the cosmic observer might have (in principle) infinite resources available (infinite number of states of the TAPE) even if only a finite portion of these resources is used at any given time. 

Regardless of what kind of information is written on the TAPE of a CO machines (e.g. discretized fields, fundamental constants, Cosmic Microwave Background radiation map, etc.)  what we want  is to be able to assign probabilities to different CO machines. For that we define another type of Turing machine, called cosmic measure (CM) machine, which takes a given CO as input and outputs its probability, i.e.
\be
\mu(\text{CO}) \equiv \text{CM}(\langle \text{CO} \rangle).
\ee
More precisely what is initially written on the TAPE of CM machine is a Turing number of CO whose probability is being calculated. Since CO is a Turing machine it may halt after a finite time (i.e. mortal CO) or run indefinitely (i.e. immortal CO), but it would always be a finite-size Turing machine and  thus would be described by a finite Turing number.  At the end of calculations the CM machine would write on its TAPE a probability of the CO whose Turing number was inputed. 

\section{Cosmic Symmetries} \label{Sec:Symmetries}

Now that we have defined a set of cosmic observers $\cal O$, or CO machines, and cosmic measures $\cal \mu$, or CM machines, we switch to the third type of machines which represent cosmological symmetries $\cal S$, or CS machines. What a given CS machine would do is take a Turing number $\langle \text{CO} \rangle$ of some CO machine as input and output $1$ for CO machines which have the same probability (with respect to the symmetry that this CS describes) and output $0$ otherwise. Note that CS machines do not say anything about probabilities of CO machines if CS's output was $0$, but they do guarantee that the probabilities of CO machines is the same if CS's outputs were $1$, i.e.
\be
\text{CS}(\langle \text{CO}_1 \rangle) = \text{CS}(\langle \text{CO}_2 \rangle)=1 \;\; \Rightarrow \;\;\mu(\text{CO}_1) = \mu(\text{CO}_2).
\ee

At this point it is not obvious what is the correct (and complete) set of CS machines or whether this set would uniquely determine the correct CM machine. On the other hand the construction of CS machines from CM machines is trivial. For example, one can use CM machine to build a set of CS$_p$ machines parametrized by probability $p$ such that CS$_p$ would output $1$ for any CO whose probability (according to CM) is $p$, i.e.
\be
\text{CS}_p (\langle \text{CO} \rangle) = \begin{cases} 
1 \;\;\;\; \text{if CM}(\langle \text{CO} \rangle) = p \\
0 \;\;\;\; \text{otherwise}.
\end{cases}
\ee
This suggests that CS machines are more fundamental and one should also think of them as building blocks for CM machines. For this reason and in order to study the cosmic symmetries without referring to a specific cosmic measure from now on we will concentrate on the analysis of CS machines. 

In fact some symmetries are common to many cosmic measures, but so far the analysis of these measure was based mostly on their phenomenological properties. In particular, it was argued that certain cut-off measures  (e.g. scale factor cut-off) lead to cosmological predictions which are in a good agreement with observations \cite{ComputerObserver, ScaleFactor}. In this section we will instead discuss the computability properties of a CS whose symmetry is common to all (global and local) cut-off measures. In particular, the cut-off measures (by construction) assign zero probability to all observers with infinite temporal extend and finite probability to all of the observers with finite temporal extend. In our language of Turing machines the CM machines computing cut-off measure would assign zero probability to all immortal CO machines and finite probability to all mortal CO machines, i.e. 
\be
\text{CS}_0 (\langle \text{CO} \rangle) = \begin{cases}  \label{Eq:CO_0}
1 \;\;\;\; \text{if CO runs forever} \\
0 \;\;\;\; \text{if CO halts}
\end{cases}
\ee
We now prove that CS machines with such symmetry are incomputable, and thus the CM machines corresponding to the cut-off measures are incomputable. 

The presented proof is closely related to the proof of the so-called blank tape halting problem. The key point to note is that for any Turing machine M and any number $x$ we can always construct another Turing machine (call it M$_x$) which would first write $x$ on the TAPE, then moves the read-write head to the leftmost position and runs as the M machine would. Then if we assume that there exist a Turing machine CS$_0$ (defined by Eq. \eqref{Eq:CO_0}) then we can run CS$_0$ with a Turing number of M$_x$ as input and it would have to output $1$ if and only if M$_x$ machine does not halt or equivalently if and only if M machine with input $x$ does not halt,
\be
\text{CS}_0(\langle \text{M}_x\rangle) = 1 \iff \text{M}_x \;\; \text{runs forever} \iff \text{M}(x) \;\; \text{runs forever}.
\ee
However, if that was the case, this would solve the halting problem which is known to be incomputable \cite{Halting}. Therefore the original assumption of the existence of CS$_0$ (which decides whether a given observer is mortal or immortal) must be incorrect. 

The above prove was based on the well-known result that the halting problem is incomputable \cite{Halting}, but for purpose of completeness we now show that this is indeed the case. In fact what we will show is that it is impossible to even decide if a given Turing machine M would halt if the input is its own Turing number $\langle \text{M} \rangle$. The proof is also by contradiction where we use the method of diagonalization that was originally introduced by Cantor to prove uncountability of real numbers and was later adopted by Turing to prove the non-computability of the halting problem. Let's assume that there exist a machine (call it HALT) which upon input $\langle \text{M} \rangle$ outputs $0$ if M would not halt upon input $\langle \text{M} \rangle$ and $1$ if M would halt upon input $\langle \text{M} \rangle$, i.e.
\be
\text{HALT}(\langle \text{M} \rangle) = \begin{cases} 
1 \;\;\;\; \text{if M(}\langle \text{M} \rangle \text{) halts} \\
0 \;\;\;\; \text{if M(}\langle \text{M} \rangle \text{) runs forever} \label{Eq:HALT}
\end{cases}
\ee
Since by assumption HALT is a valid machine we can construct another machine (call it DIAG) which does the following:
\begin{align}
 & y = \text{HALT}(x);\notag \\
& \text{if}  \;\; y=0 \;\; \text{then} \notag \\
& \;\;\;\;\;\;\;\;\;\;  \text{halt;}\notag \\
 & \text{else}  \notag \\
& \;\;\;\;\;\;\;\;\;\;   \text{loop forever;} \notag \\
& \text{end if} \label{Eq:DIAG}
\end{align}
Then upon input of its Turing number $\langle$DIAG$\rangle$ into DIAG  machine we have two options both of which are unacceptable:
\begin{enumerate}
\item If HALT($\langle$DIAG$\rangle$)=0, then DIAG($\langle$ DIAG $\rangle$) must not halt according to \eqref{Eq:HALT}, and must halt according to \eqref{Eq:DIAG}.
\item If HALT($\langle$DIAG$\rangle$)=1, then DIAG($\langle$ DIAG $\rangle$) must halt according to \eqref{Eq:HALT}, and must loop forever according to \eqref{Eq:DIAG}.
\end{enumerate}
Thus a contradiction is reached which proves that the machine HALT does not exist. And as we have already shown this implies that CS$_0$ machine does not exist.

This is an important result as it tells us something about  measures most often analyzed in context of  eternal inflation namely the cut-off measures. Indeed for any finite cut-off the fraction of infinite word-lines (or infinite stories in the language of Ref. \cite{GuthVanchurin}) is not only small, but is exactly zero. This implies that all of the cut-off measures share the symmetry of CS$_0$ that puts all of the immortal observers into the same equivalence class. But since the CS$_0$ machine does not exist, all of the cut-off measures (global or local) must be incomputable.

There are two possible ways out of the non-computability conclusion which should be explored further. One possibility is to postulate a new scale $T$ and to modify the set of CO machines by requiring that all of CO machines halt after a finite number of steps $T$. This would save all of cut-off measure, but it is unclear what this new scale should be. Another possibility is to only allow measures which do not share CS$_0$ symmetry and thus are not forbidden by the computability analysis presented in this section. \footnote{See for example Refs.\cite{Markov}, \cite{Vanchurin} and \cite{Continuum} for our recent proposals based on the theory of Markov processes, the theory of dynamical systems and measure theory respectively.}

\section{Summary}\label{Sec:Discussion}

We now summaries the main results of the paper:

\begin{enumerate}
\item We described the measure problem as the problem of defining a triplet $({\cal O}, {\cal S}, \mu)$ where $\cal O$ is the set of cosmic observers/observations, $\cal S$ is the group of cosmic symmetries and $\mu$ is cosmic measure which may or may not be unique depending on both ${\cal O}$ and ${\cal S}$. We claim that the triplet is what needed to make logical inferences about universe and is what we call in general cosmic logic. 
\item We proposed a simple computational model of cosmic logic with all three components described by Turing machines: $\cal O$ is a set of CO machines, $\cal S$ is a set of CS machines, and $\mu$ is a CM machine. CO machines start with a blank TAPE and describe the temporal development of the state of cosmic observer which may (i.e. moral CO) or may not (i.e. immortal CO) halt in finite time. CS machines take CO's Turing number as input and output one if the CO belongs to the equivalence class described by the cosmic symmetry of the CS (and zero otherwise). CM machines also take CO's Turing number as input and output its probability. 
\item We showed that CS machines are more fundamental and proposed to use them as building blocks in constructing CM machines. An important example of CS is a machine which outputs one for all immortal CO machines and zero for all mortal CO machines. We proved that such a CS machine does not exist and used it to argue that all of the global/local cut-off measures are incomputable. The proof is based on the well-known non-computability of the so-called halting problem \cite{Halting}.

\item We argued that the description of cosmic observers in terms of Turing machines necessarily splits the state into degrees of freedom of the REGISTER and of degrees of freedom of the TAPE. This division is unusual from the point of view of physical systems where all of the degrees of freedom are treated on equal footings. A natural (although counterintuitive) choice in context of cosmic observers was to associate the external degrees of freedom with the REGISTER and the internal degrees of freedom with the TAPE. 

On a (lot) more speculative side the splitting of degrees of freedom (between the REGISTER and the TAPE) might be relevant to the so-called mind-body problem \cite{MindBody}. In particular, the description of observers with Turing machines supports the dualism approach \cite{Dualism} where there is a clear distinction between the state of mind (i.e. the TAPE in case of cosmic observers) and the state of matter (i.e. the REGISTER in case of cosmic observers). 
 \end{enumerate}

{\it Acknowledgments.} The work was supported in part by Templeton Foundation and Foundational Questions Institute (FQXi).

\appendix

\section{Discrete Universe}\label{Sec:DiscreteUniverse}

In this appendix we describe a procedure of discretizing arbitrary  (Lorentzian) manifold  $\cal M$  and fields $\varphi$  (i.e. sections over  some fiber ${\cal F}$ with base space $\cal M$) where both $\cal M$  and $\varphi$ need not satisfy any particular equations of motion (i.e. we make no assumption of general relativity, standard model, etc.) For definiteness we assume that both the manifold $\cal M$ and  the fiber ${\cal F}$  are infinite and thus will be discretized with a countable number of points.  This assumption is not essential and one could carry on the discretization procedure for finite $\cal M$ and/or finite ${\cal F}$. 

The discretization is done in three steps: first we discretize $\cal M$, then we discretize $\cal F$ and finally we discretize $\varphi$. To discretize the manifold $\cal M$ we construct a set of vertices defined by the following map 
\be
V:\mathbb{N} \rightarrow {\cal M} \label{eq:V}
\ee
which is sufficiently dense in a sense that
\be
\forall x \in {\cal M}\; \exists n \in \mathbb{N} \; \text{such that} \; \| d(x,V_n) \| < 1
\ee
but not too dense in  a sense that 
\be
\forall n \in \mathbb{N} \; \exists x \in {\cal M} \; \text{such that} \; \inf_{i\in \mathbb{N}\setminus \{ n\}} \| d(x,V_n) \| > 1
\ee
where $d(x,y)$ is the metric distance between point $x$ and $y$ (in for example Planck units) and by $\| \cdot \|$ we denote absolute value. Then we define a set of directed edges connecting the vertices in chronological order only if the distance between vertices is less than unity i.e., 
\be
E(V_i, V_j) = \begin{cases} \label{eq:E}
1 \;\; \text{if $V_j \in I^+(V_i)$ and $\| d(V_i,V_j) \| < 1$} \\
0 \;\; \text{otherwise}
\end{cases}
\ee
where $I^+(x)$ is the chronological future of point $x$. 

Similarly, to discretize the fiber ${\cal F}$ we define 
\be
F: \mathbb{N} \rightarrow {\cal F}  \label{eq:F}
\ee
which is sufficiently dense in a sense that
\be
\forall f \in {\cal F}\;  \exists n \in \mathbb{N}  \; \text{such that} \; D(f,F_n) < 1
\ee
but not too dense in a sense that 
\be
\forall n \in \mathbb{N} \;  \exists f \in {\cal F}  \; \text{such that} \; \inf_{i\in \mathbb{N}\setminus \{ n\}} D(f,F_n) > 1
\ee
where $D(x,y)$ is some suitably defined metric on configuration space such that point which are separated by less than unity would be experimentally indistinguishable.\footnote{It might be that only a finite number $N$ of field values are experimentally distinguishable and then the set of natural numbers $\mathbb{N}$ may be replaced with a finite set $\{1, 2, ..., N\}$ in the definition for $F$.} 

And finally, given any realization of fields $\varphi$ in space-time $\cal M$, we can define discretized fields with
\be
\Phi: \mathbb{N} \rightarrow  \mathbb{N}
\ee
as a map defined by  
\be
D(\varphi(V_i), F_{\Phi_i}) \equiv \inf_{m \in\{1,...,N\}} D(\varphi(V_i), F_m) \}.
\ee
for all $i \in \mathbb{N}$. Thus given a triplet $\{{\cal M}, {\cal F}, \varphi \}$ representing the space-time with fields one can always define discretized space-time and discretized fields by $\{V, E, F, \Phi \}$.


\begin{thebibliography}{10}


\bibitem{EternalInflation}
P.~J.~Steinhardt, ``Natural inflation,'' in: The Very Early Universe, Proceedings of the Nuffield Workshop, Cambridge 1982, pp. 251-266;
\bibitem{EternalInflation2}
A.D. Linde,  ``Nonsingular Regenerating Inflationary Universe,'' Print-82-0554, Cambridge University preprint, 1982
\bibitem{EternalInflation3}
A.~Vilenkin, ``The Birth Of Inflationary Universes,'' Phys.~Rev.~D {\bf 27}, 2848 (1983)


 \bibitem{Vilenkin}
  V.~Vanchurin, A.~Vilenkin and S.~Winitzki, ``Predictability crisis in inflationary cosmology and its resolution,'' Phys.\ Rev.\  {\bf D61}, 083507 (2000), [gr-qc/9905097] 

\bibitem{GuthVanchurin}
  A.~H.~Guth and V.~Vanchurin, ``Eternal Inflation, Global Time Cutoff Measures, and a Probability Paradox,'' arXiv:1108.0665 [hep-th]
  
\bibitem{Vanchurin}
   V.~Vanchurin,``Dynamical systems of eternal inflation: a possible solution to the problems of entropy, measure, observables and initial conditions,''
  Phys.\ Rev.\ D {\bf 86}, 043502 (2012)
  [arXiv:1204.1055 [hep-th]].

\bibitem{Continuum} V.~Vanchurin,
  ``Continuum of discrete trajectories in eternal inflation,''
  Phys.\ Rev.\ D {\bf 91}, no. 2, 023511 (2015)
  [arXiv:1408.4854 [hep-th]].

\bibitem{Tree}
D.~Harlow, S.~H.~Shenker, D.~Stanford and L.~Susskind, ``Eternal Symmetree,'' arXiv:1110.0496 [hep-th];

\bibitem{Haar}
Alfsen, E.M. (1963), ``A simplified constructive proof of existence and uniqueness of Haar measure'', Math. Scand. 12: 106Ð116

\bibitem{GuthVanchurin2}
R.~Bousso, B.~Freivogel, S.~Leichenauer and V.~Rosenhaus, ``Eternal inflation predicts that time will end,'' Phys.\ Rev.\ D {\bf 83}, 023525 (2011)  [arXiv:1009.4698 [hep-th]]

\bibitem{GuthVanchurin3}
K.~D.~Olum, ``Is there any coherent measure for eternal inflation?,'' arXiv:1202.3376 [hep-th]


\bibitem{GuthVanchurin4}
M.~Noorbala and V.~Vanchurin, ``Geocentric cosmology: A New look at the measure problem,'' [arXiv:1006.4148 [hep-th]].
   
   
   
 \bibitem{Youngness}
A.~D.~Linde, D.~A.~Linde and A.~Mezhlumian,
  ``Do we live in the center of the world?,''
  Phys.\ Lett.\ B {\bf 345}, 203 (1995)
  [hep-th/9411111];
  
   \bibitem{Youngness2}
   H.~Guth, ``Inflation and eternal inflation,'' Phys.\ Rept.\  {\bf 333}, 555 (2000), [arXiv:astro-ph/0002156];
   
    \bibitem{Youngness3}
   M.~Tegmark, ``What does inflation really predict?,'' JCAP {\bf 0504}, 001 (2005), [astro-ph/0410281];




\bibitem{Boltzmann}
L.~Dyson, M.~Kleban and L.~Susskind, ``Disturbing implications of a cosmological constant,'' JHEP {\bf 0210}, 011 (2002), [arXiv:hep-th/0208013]

\bibitem{Boltzmann2}
A.~Albrecht and L.~Sorbo, ``Can the universe afford inflation?,'' Phys.\ Rev.\  D {\bf 70}, 063528 (2004), [arXiv:hep-th/0405270]

\bibitem{Boltzmann3}
D.~N.~Page, ``The Lifetime of the universe,''  J.\ Korean Phys.\ Soc.\  {\bf 49}, 711 (2006), [arXiv:hep-th/0510003]

  
\bibitem{Halting}
 A.~Turing, ``On computable numbers, with an application to the Entscheidungsproblem'', Proceedings of the London Mathematical Society, Series 2, 42 (1936-7), pp 230Ð265.

\bibitem{NumericalSearch} 
  V.~Vanchurin,
  ``Numerical search for fundamental theory,''
  Phys.\ Rev.\ D {\bf 77}, 043503 (2008)
  [hep-th/0701147].


\bibitem{ComputerObserver}
 A.~De Simone, A.~H.~Guth, A.~D.~Linde, M.~Noorbala, M.~P.~Salem and A.~Vilenkin, ``Boltzmann brains and the scale-factor cutoff measure of the multiverse,'' Phys.\ Rev.\  D {\bf 82}, 063520 (2010), [arXiv:0808.3778 [hep-th]]


   
\bibitem{ScaleFactor}
 A.~De Simone, A.~H.~Guth, M.~P.~Salem and A.~Vilenkin, ``Predicting the cosmological constant with the scale-factor cutoff measure,'' Phys.\ Rev.\ D {\bf 78}, 063520 (2008),  [arXiv:0805.2173 [hep-th]];   


\bibitem{Markov} 
  V.~Vanchurin,
  ``Geodesic measures of the landscape,''
  Phys.\ Rev.\ D {\bf 75}, 023524 (2007)
  [hep-th/0612215].
  


\bibitem{MindBody}
R.~M.~Young, ``The mindÐbody problem'', In RC Olby, GN Cantor, JR Christie, MJS Hodges, eds. Companion to the History of Modern Science (Paperback reprint of Routledge 1990 ed.). Taylor and Francis. pp. 702Ð11. (1996)


\bibitem{Dualism} W~D.~Hart, ``Dualism'', in A Companion to the Philosophy of Mind, ed. Samuel Guttenplan, Oxford: Blackwell, pp. 265-7.  (1996)

\end{thebibliography}
\end{document}